# SOAP3-dp: Fast, Accurate and Sensitive GPU-based Short Read Aligner

Ruibang Luo[1,*], Thomas Wong[1,*], Jianqiao Zhu[1,4,*], Chi-Man Liu[1,*], Edward Wu[1], Lap-Kei Lee[1], Haoxiang Lin[2], Wenjuan Zhu[2], David W. Cheung[1], Hing-Fung Ting[1], Siu-Ming Yiu[1], Chang Yu[2], Yingrui Li[2], Ruiqiang Li[3], Tak-Wah Lam[1]

[1] HKU-BGI Bioinformatics Algorithms and Core Technology Research Laboratory & Department of Computer Science, University of Hong Kong, Hong Kong.
[2] BGI-Shenzhen, Shenzhen, Guangdong 518083, China.
[3] Peking-Tsinghua Center for Life Sciences, Biodynamic Optical Imaging Center and School of Life Sciences, Peking University, China.
[4] Department of Computer Sciences, University of Wisconsin-Madison, USA.
*These authors contributed equally to this work.

Correspondence should be addressed to T. L. (twlam@cs.hku.hk), Ruiq. L. (lirq@pku.edu.cn) or Y. L. (liyr@genomics.cn).

| | |
|---|---|
| Ruibang Luo | rbluo@cs.hku.hk |
| Thomas Wong | thomaskf@gmail.com |
| Jianqiao Zhu | jianqiao@cs.wisc.edu |
| Chi-Man Liu | cmliu@cs.hku.hk |
| Edward Wu | edwardmkwu@gmail.com |
| Lap-Kei Lee | lklee@cs.hku.hk |
| Haoxiang Lin | linhaoxiang@genomics.cn |
| Wenjuan Zhu | zhuwenjuan@genomics.cn |
| David W. Cheung | dcheung@cs.hku.hk |
| Hing-Fung Ting | hfting@cs.hku.hk |
| Siu-Ming Yiu | smyiu@cs.hku.hk |
| Chang Yu | yuchang@genomics.cn |
| Yingrui Li | liyr@genomics.cn |
| Ruiqiang Li | lirq@pku.edu.cn |
| Tak-Wah Lam | twlam@cs.hku.hk |

# Abstract:

To tackle the exponentially increasing throughput of Next-Generation Sequencing (NGS), most of the existing short-read aligners can be configured to favor speed in trade of accuracy and sensitivity. SOAP3-dp, through leveraging the computational power of both CPU and GPU with optimized algorithms, delivers high speed and sensitivity simultaneously. Compared with widely adopted aligners including BWA, Bowtie2, SeqAlto, CUSHAW2, GEM and GPU-based aligners BarraCUDA and CUSHAW, SOAP3-dp was found to be two to tens of times faster, while maintaining the highest sensitivity and lowest false discovery rate (FDR) on Illumina reads with different lengths. Transcending its predecessor SOAP3, which does not allow gapped alignment, SOAP3-dp by default tolerates alignment similarity as low as 60%. Real data evaluation using human genome demonstrates SOAP3-dp's power to enable more authentic variants and longer Indels to be discovered. Fosmid sequencing shows a 9.1% FDR on newly discovered deletions. SOAP3-dp natively supports BAM file format and provides the same scoring scheme as BWA, which enables it to be integrated into existing analysis pipelines. SOAP3-dp has been deployed on Amazon-EC2, NIH-Biowulf and Tianhe-1A.

# Maintext:

**Introduction:**
With the rapid advancement of Next-Generation Sequencing technologies, modern sequencers like Illumina HiSeq 2500 can sequence a human genome into 600 million pairs of reads of 100bp in length (total 120 Gigabases) in merely 27 hours.  The cost is also decreasing fast. By 2013 year's end, sequencing a human genome is projected to cost less than $1,000.  Bioinformatics research using sequencing data often starts with aligning the data onto a reference genome, followed by various downstream analyses. Alignment is computationally intensive; the 1000 genomes pilot paper[1] published in 2010 reported that a 1192-processor cluster was used to align the reads using MAQ[2]. This kind of computing resources is not available to most laboratories, let alone clinical settings. Although considerable advances have been made on new aligners, alignment still remains a bottleneck in bioinformatics analyses. Thus, ultra-fast alignment tools without relying on extensive computing resources are needed.

There are quite a few software tools for aligning short reads onto a reference genome. The more popular ones include MAQ, Bowtie[3], BWA[4] and SOAP2[5]. The faster ones[3-5] index the reference genome using the Burrows-Wheeler Transform (BWT), which is efficient for aligning short reads with limited mismatches, but inefficient for alignment with gaps.  These tools (running on a quad-core processor) take tens of hours to align 120 Gigabases with limited (or even none for Bowtie and SOAP2) gapped alignment found. Alignment gaps can result from insertions and deletions (Indels), which are thought to comprise over 20% of genetic variations[6] and contribute to human traits[7]. Hence, a successor is expected to be faster and more sensitive to gaps.

SeqAlto[8], CUSHAW2[9], and GEM[10] were published recently. SeqAlto is a hash-based aligner that improves an earlier hash-based aligner SNAP[11] (reported to have relatively poor sensitivity for real reads and provide no mapping quality[8]) using additional global and local alignments. SeqAlto is slower than SNAP, yet SeqAlto is still faster than the BWT-based aligners except Bowtie2. CUSHAW2[9] uses the seed-and-extend approach and maximal-exact-match seeds to enable gapped alignment of long reads. GEM mapper leverages string matching with filtration to search the alignment space more efficiently[10]. GEM is faster than comparable state-of-the-art aligners. Yet it does not provide PHRED[12] compliant mapping quality score; this saves some tedious computation, but prohibiting it from integrating into existing analysis pipelines.

Nowadays, general-purpose computing on graphics processing units (GPUs) is becoming popular. A GPU is a piece of low-cost hardware providing massive parallelism but with limited memory and restricted usage. A number of GPU-based bioinformatics tools have emerged last year[13]. CUSHAW is the first to introduce a complete alignment pipeline utilizing GPU power for paired-end short reads (note that CUSHAW2, mentioned above, is CPU-based). BarraCUDA[14] implements BWA to align reads in parallel on a GPU; limited by the branch and bound trie algorithm that requires extensive decisions making, BarraCUDA works sub-optimally on GPU and gains a 4-time boost than a single-thread BWA. SOAP3[15] successfully exploits the massive parallelism of a GPU with tailor-made GPU-BWT and read-characteristics sensitive load balancing to effectively align short reads. Albeit not supporting gapped alignment, which makes it unsuitable for production, it is to date the most competitive aligner for ungapped alignment.

Here we present a GPU-based alignment software SOAP3-dp that allows multiple mismatches and gaps, which is well suited for production environments (real data alignments) than the predecessor SOAP3. A simple approach to extend mismatch alignment to gapped alignment is to first identify candidate regions by exact or mismatch alignment of short substrings (seeds) in the reads, then use dynamic programming to perform a detailed alignment of the read to the regions. Such an approach has been widely used (e.g., Bowtie2). The bottleneck occurs as substring alignment often results in a large number of candidates, especially when mismatches are allowed. As a result, reads with too many candidates are often ignored due to time constraint. On the other hand, the parallelism of GPU apparently would allow many candidates to be verified in parallel; yet dynamic programming is memory consuming, and the limited-memory of GPU becomes a prohibiting factor to fully utilize the parallelism. SOAP3-dp gives a pragmatic realization of this approach (Figure 1, Methods). By exploiting compressed indexing and memory-optimizing dynamic programming on a GPU, SOAP3-dp can efficiently tackle a large number of candidates in parallel, and thus can examine gapped alignments extensively and achieve a drastic improvement in both speed and sensitivity over other tools.

See the Methods about the design of the dynamic programming which attempts to minimize the memory usage for each candidate so as to let a GPU to handle hundreds of candidates in parallel while using limited shared memory. We also show the details of SOAP3-dp's intricate engineering solution to finding the optimal way to align different reads using either the CPU or the GPU.

**Experiments & Performance:**
We compared SOAP3-dp to other short-read alignment software in terms of the speed, sensitivity and accuracy. We used both real and simulated Illumina data. Furthermore, we tested out SOAP3-dp's alignment quality for variant calling using real data. In particular, 41-fold of 100bp (PE100) and 77-fold of 150bp (PE150) Illumina paired-end reads of YH[16] samples have been generated (Supplementary Table 1) for the testing for variant calling.

**Alignment Performance.** We first used real data to test SOAP3-dp with BWA (BWA-SW[17] for SE reads), Bowtie2, SeqAlto, CUSHAW2, GEM, BarraCUDA, and CUSHAW. The aim was to compare the time and alignment rate when each runs in the default setting. Next, to assess the accuracy and sensitivity, we used simulated reads whose correct alignments were known. We then considered more detailed comparison with the software running in different settings. We also attempted to compare SOAP3-dp against its predecessor SOAP3.

In our experiments, we assume that input reads are plain text instead of in gzip-compressed format. This is because GEM (to the date of paper submission) does not accept gzip-compressed FASTQ file. All other software can handle compressed input, which is getting common nowadays. Regarding output format, we require all software to use SAM format, which is mandatory for downstream analysis software including GATK[18] and SAMTOOLS[19]. All software except GEM can output directly in SAM format; GEM first outputs in a simple format and then takes an extra step to convert to SAM format. To test GEM's efficiency when using a simple format, we include a comparison to SOAP3-dp also using a simple format (see the Remark section).

**Real data.** We used three real datasets for benchmarking of alignment performance: (1) a lane (122.43M reads) from PE100, and (2) a lane (374.87M reads) from PE150, and (3) SRR211279 (25.23M 100bp paired-end reads generated by Illumina GAIIx) from the Washington University Genome Sequencing Center. We tested SOAP3-dp and seven other aligners (CPU-based: BWA, Bowite2, SeqAlto, CUSHAW2, GEM; GPU-based: Barracuda, and CUSHAW; see Supplementary Note for receipts), all using 4 CPU threads and one GPU device (for GPU aligner). As shown in Table 1, SOAP3-dp is much faster than all others (Supplementary Tables 2-10 for more details). It is at least 3.5 times faster than GEM, and 7 to 15 times faster than the other six. SOAP3-dp also gave better alignment rate consistently. SeqAlto comes closest, aligning 0.48% to 3.6% less reads than SOAP3-dp, and the others are in the range of 2% to 8% less than SOAP3-dp. Notice that except SOAP3-dp, aligners usually have an obvious drop in alignment rate for longer reads (dataset 2). The two GPU-based aligners, Barracuda and CUSHAW, are relatively primitive in optimizing GPU's utilization and overheads, and their performance was dominated by new CPU-based aligners like GEM and Bowtie2. For SOAP3, its alignment rate, as expected, is much poorer than SOAP3-dp and the others (due to lack of gapped alignment); furthermore, SOAP3 is slower than SOAP3-dp for 100bp reads. We did not include Barracuda, CUSHAW and SOAP3 for further experiments.

**Simulated data**. To assess the accuracy and sensitivity of SOAP3-dp, we used the short read simulator Mason[20] to obtain 5 sets of 6M Illumina-style paired-end (PE) reads with 500bp insert size from GRCh37 major build, with length ranging from 50-250bp.

Notably, Bowtie2, SeqAlto and GEM were designed with switches to favor speed at the expense of accuracy and sensitivity. We applied "very-fast", "sensitive", and "very-sensitive" switches to Bowtie2, "fast (-f)" to SeqAlto and "fast adaptive (--fast-mapping)", "fastest (--fast-mapping=0)" to GEM. For SOAP3-dp, we tested three versions whose indices are based on 1/4 sampled, 1/2 sampled and full suffix array (SA), respectively. Different sized SAs still deliver identical alignment results, but a smaller one consumes less memory at the expense of slightly longer alignment time. All parameters of SOAP3-dp and SOAP3 remained as default (for a 100bp read, one gap up to 68bp, to 23 one-bp gaps) while parameters for other aligners were set to favor different read types and lengths as suggested by previous studies. Detailed command lines and descriptions of critical parameters were summarized as receipts in the Supplementary Note. In total, 16 sets of programs and parameters were compared.

In all datasets of simulated reads, SOAP3-dp gives consistent advantage. It is faster and simultaneously has higher sensitivity and lower FDR over all other tools (Table 2 for 100bp PE, Supplementary Tables 11-14 for all other simulated datasets, Figure 2). For 100bp reads, SOAP3-dp with full SA takes 132 seconds to align 6M read pairs, and it is 2.26 to 12.63 times faster than the others (others using the fastest switches). SOAP3-dp's sensitivity is 99.96%, which is 0.13 - 0.85% higher than the others (others using the sensitive switches), and SOAP3-dp's FDR is 0.34%, which is lower than the others by 0.13 - 0.85%. Apparently the simulated data is easier to align than the real data due to recombination hotspots with intensified variants in real genome [6]. SOAP3-dp consumes more memory (9.3, 11.9, 17.2GB for 1/4, 1/2 and full SA in average, respectively) than other software; Bowtie2 has the least (3.5GB). Nevertheless, workstations and servers nowadays are equipped with at least 16GB or even 32GB of memory; SOAP3-dp is designed to take advantage of the available memory to achieve speed.

Mapping quality score is mandatory for most of the popular downstream analysis tools such as GATK and SAMTOOLS. SOAP3-dp uses the same scoring scheme as BWA so as to make its alignment results compatible to the expectations of existing analysis tools. As shown in Figure 3 (Supplementary Figure 1a,b), BWA, Bowtie2, SeqAlto, BarraCUDA, CUSHAW2 and SOAP3-dp provide mapping quality scores that can differentiate different alignments properly, while GEM's scores are too rough, and too many incorrectly aligned reads are given high quality scores, which makes it unsuitable for downstream analysis.

**Remarks.** Note that GEM can output in a simple format to save time. When compared to SOAP3-dp in its own simple format, GEM and SOAP3-dp can both save about half of their running time; for the 6M simulated paired-end data of length 100bp, the alignment time of GEM and SOAP3-dp is reduced to 90 seconds and 38 seconds, respectively. Downstream programs for variants calling, if redesigned to utilize these specific formats, could save time.

The simulated dataset is relatively small (6M read pairs), thus when using a large SA, the index loading time of SOAP3-dp dominated the total elapsed time. Considering only the alignment time (time consumption after index loading, including input of raw reads and output of alignment results), SOAP3-dp using the full and 1/2 SA is 12 and 9 seconds faster than 1/4 SA, thus for large real datasets, 1/2 and full SAs are suggested if memory permits.

Different generations of GPU device differ in speed. We compared the performance of SOAP3-dp between the latest GPU "GTX680" and a previous generation "Tesla C2070" using simulated datasets. The alignment time extended about 10% for each dataset (Figure 4) using "Tesla C2070". Furthermore, a large real set of 150bp paired-end reads was used. The alignment using the "GTX680" consumed 6,835 seconds, which is 4,658 seconds (1.68 times) faster than "Tesla C2070".

**Variant Calling Performance:**
Next, we considered SOAP3-dp's alignment quality for variant calling. The full sets of both PE100 and PE150 were aligned using SOAP3-dp. We used the widely adopted BWA as benchmark. With the alignment results, variants were called using GATK's UnifiedGenotyper[21] and filtered by VariantRecalibrators, with and without GATK's local realignment (see Methods). Before we detail the results on variant detection, it is worth mentioning that BWA, even running in a slower mode to allow a longer gap (one gap up to 50bp, without "-m" parameter to allow hit entries higher than 2 million due to out of memory error), still cannot catch up the sensitivity of SOAP3-dp in default setting (for a 100bp read, one gap up to 68bp, to 23 one-bp gaps, Table 3).

SOAP3-dp's better sensitivity is due to its ability of extensive gapped alignment; the extra reads aligned are crucial for variant detection (in particular, Indels). SOAP3-dp allowed 2.4% and 4.0% more SNPs than BWA, and 6.1% and 9.8% more Indels for the two datasets PE100 and PE150, respectively. Intuitively, longer reads are more favorable for variant detection; this is indeed reflected in SOAP3-dp's performance, but not for BWA (Table 3). We further checked the SNPs detected against dbSNP v135 (an archive of SNPs validated by previous studies); SOAP3-dp has notably 2.1% and 3.6% more SNPs found in dbSNP, confirming a higher sensitivity.

SOAP3-dp allowed more Indels to be detected than BWA, especially more Indels longer than 20bp (Figures 5a, b, Supplementary Figure 2). To validate the novel Indels detected, we randomly selected 50 deletions that SOAP3-dp exclusively detected and are not yet archived in dbSNP v135, and verified them using Fosmid sequencing (see Methods). The Fosmids were sequenced, assembled and then aligned to the reference genome. The 50 deletions were covered by 460 Fosmid sequences. The findings are as follows: 6 deletions were inconclusive due to insufficient coverage of Fosmid sequences, 40 deletions were validated, and 4 rejected, revealing a FDR of 9.1% (Supplementary Tables 15, 16). SOAP3-dp's ability to allow long gaps without speed penalty provides an unprecedented opportunity to come up with a more comprehensive Indel identification in large-scale genome studies.

With SOAP3-dp's ability to authentically align more reads, more multi-nucleotide polymorphisms (MNP) among the whole genome were identified (Supplementary Figure 3). Notably, GATK's local realignment can eliminate inauthentic alignments and rescue true variants. For SOAP3-dp, the number of MNP increased by 4.1% after realignment; yet for BWA, the number decreased by 6.3% (Supplementary Table 17), indicating that SOAP3-dp initially provided much more reliable alignments and led to more accurate variant calling.

**Discussion:**
SOAP3-dp has been successfully deployed on Amazon EC2, NIH BioWulf and Tianhe-1A computing-cloud. On Amazon EC2, users can access SOAP3-dp's program and a testing dataset by mounting EBS snapshot "snap-154f1c54" named "SOAP3-dp" while creating a GPU instance (Supplementary Note). To test out SOAP3-dp's performance on Amazon EC2, we selected 10 Illumina PE datasets from 1000 genomes project, comprising 131.44Gbp of raw reads (43.8-fold). The 10 datasets were distributed to the two available Tesla M2050 GPU cards (see Methods and Supplementary Note. Notably, Tesla M2050 is slower than the GTX680 and Tesla C2070 we have used for real and simulated data evaluation) with one copy of the index shared in host memory. Using default parameters and BAM output, the alignment finished in 3.8 hours, yielding a total cost of $7.98, or $0.061 per Gbp reads aligned.

For users' convenience, SOAP3-dp separates the output of reads into three categories: 1) alignments involve gaps and extensive mismatches; 2) few mismatches only and 3) improperly paired or unaligned (file suffix "dpout", "gout", and "unpair" respectively). The separate file scheme fits well with the production environments, where files could be sorted separately in parallel and then merged together, which saves time than sorting a single SAM file. The files could also be concatenated by SAMTOOLS easily.

SOAP3-dp does not enforce a maximum read length. However, read length longer than 500bp is not recommended while the current version of SOAP3-dp is tailor-made for Illumina reads. A version for longer 454 reads and Ion Torrent reads without performance degradation is our next task.

Overall, SOAP3-dp is an efficient alignment tool that targets the future of genome analysis where reads are longer and the volume is larger. SOAP3-dp is much faster than existing tools while retaining the ability to align more reads correctly. To be flexible, SOAP3-dp outputs both SAM and BAM formats that are compatible with most downstream analysis tools. SOAP3-dp is a free and open-source alignment tool available at http://www.cs.hku.hk/2bwt-tools/soap3-dp/.

# Acknowledgements

Tak-Wah Lam was partially supported by RGC General Research Fund 10612042. Ruibang Luo was supported by Hong Kong ITF Grant GHP/011/12.

# Author Contributions

Ruib. L., T. W., J. Z. and C. L. contributed equally to this work. T. L. led the project and contributed the main ideas. Ruiq. L., Y. L., D. C., H. T. and S. Y. co-managed the project. J. Z., T. W., C. L., Ruib. L., E. W., L. L. wrote the program. Ruib. L. and T. W. designed the analysis. Ruib. L. and T. W. conducted the simulation analysis. Ruib. L., H. L. and W. Z. conducted the real data analysis. C. Y. assisted in deploying the program. T. L., Ruib. L. and T. W. wrote the paper.

# Competing Interests

The authors have declared that no competing interests exist.

# Tables

Table 1. Benchmarking using real reads. The percentage of reads aligned and time consumption aligners other than SOAP3-dp are recorded as the difference and ratio based on SOAP3-dp's figures. The '%' column represents 'Properly paired' for PE reads. "Peak Mem." represents the peak memory consumption during alignment.

| Dataset | Volume (Gbp) | GPU based ||||||||||||| CPU based |||||||||||||
|---|---|---|---|---|---|---|---|---|---|---|---|---|---|---|---|---|---|---|---|---|---|---|---|---|---|
| | | SOAP3-dp (Full SA) ||| SOAP3 ||| BarraCUDA ||| CUSHAW ||| BWA ||| Bowtie2 ||| SeqAlto ||| GEM ||| CUSHAW2 |||
| | | % | Time (s) | Peak Mem. (G) | % | Time (fold) | Peak Mem. (G) | % | Time (fold) | Peak Mem. (G) | % | Time (fold) | Peak Mem. (G) | % | Time (fold) | Peak Mem. (G) | % | Time (fold) | Peak Mem. (G) | % | Time (fold) | Peak Mem. (G) | % | Time (fold) | Peak Mem. (G) |
| realYHPE100 | 12.24 | 98.12% | 1,079 | 19 | -10.60% | 2.63 | 21.1 | -6.39% | 14.03 | 4.1 | -3.13% | 46.79 | 2.7 | -4.14% | 16.03 | 4.9 | -3.87% | 12.04 | 3.5 | -0.48% | 14.25 | 7.2 | -2.86% | 3.54 | 4.5 | -1.39% | 12.09 | 3.6 |
| realYHPE150 | 56.23 | 97.16% | 6,835 | 19.6 | -28.33% | 0.64 | 22.7 | -10.99% | 7.22 | 4.3 | -10.11% | 24.57 | 3.1 | -8.05% | 15.26 | 5.0 | -7.45% | 7.82 | 3.5 | -3.65% | 17.69 | 7.2 | -5.85% | 3.76 | 5 | -6.47% | 8.04 | 3.6 |
| SRR211279 | 5.07 | 97.21% | 439 | 18.4 | -6.83% | 1.05 | 20.8 | -5.08% | 13.42 | 4.1 | -2.22% | 54.22 | 2.1 | -2.81% | 16.27 | 4.9 | -0.69% | 12.00 | 3.5 | -0.48% | 17.29 | 7.2 | -2.44% | 4.11 | 4.6 | -2.14% | 12.03 | 3.6 |

Table 2. Comparison on 16 sets of programs and parameters using 100bp paired-end simulated reads.

| 6M 100bp Paired-end reads, 1.2Gbp bases. 500bp insert size, 25bp standard deviation. ||| GPU ||||| CPU |||||||||||
|---|---|---|---|---|---|---|---|---|---|---|---|---|---|---|---|---|---|---|
| | | | SOAP3-dp[1] (1/4 SA) | SOAP3-dp[1] (1/2 SA) | SOAP3-dp[1] (Full SA) | SOAP3-dp (Full SA, Binary) | SOAP3-dp (Full SA, Succinct) | SOAP3 | Bowtie2 (Sensititve) | Bowtie2 (Very-Sensititve) | Bowtie2 (Very-fast) | BWA[2] | SeqAlto | SeqAlto (Fast alignement) | CUSHAW2 | GEM[3] | GEM[3] (Fast Mapping: adaptive) | GEM[3] (Fast Mapping: 0) |
| Configuration | CPU (thread: core i7-3930k) | | 4 | 4 | 4 | 4 | 4 | 4 | 4 | 4 | 4 | 4 | 4 | 4 | 4 | 4 | 4 | 4 |
| | GPU (device: GTX680) | | 1 | 1 | 1 | 1 | 1 | 1 | 0 | 0 | 0 | 0 | 0 | 0 | 0 | 0 | 0 | 0 |
| Computational Resources | Total Elapsed | sec. | 132 | 137 | 162 | 111 | 112 | 132 | 966 | 1974 | 672 | 1154 | 495 | 379 | 1303 | 416 | 446 | 298 |
| | | Fold | - | 1.04 | 1.23 | 0.84 | 0.85 | 1.00 | 7.32 | 14.95 | 5.09 | 8.74 | 3.75 | 2.87 | 9.87 | 3.15 | 3.38 | 2.26 |
| | Loading Index[4] | sec. | 32 | 46 | 74 | 74 | 74 | 74 | 38 | 38 | 38 | 53+1+1 | 96 | 96 | 40 | 40+1 | 40+1 | 40+1 |
| | Alignment[5] | sec. | 100 | 91 | 88 | 37 | 38 | 58 | 928 | 1936 | 634 | 370+369+360 | 399 | 283 | 1263 | 199+176 | 238+167 | 90+167 |
| | | Fold | - | 0.91 | 0.88 | 0.37 | 0.38 | 0.58 | 9.28 | 19.36 | 6.34 | 10.99 | 3.99 | 2.83 | 12.63 | 3.75 | 4.05 | 2.57 |
| | Avg. Memory | GB | 9.3 | 11.9 | 17.2 | 17.2 | 17.2 | 17.3 | 3.3 | 3.3 | 3.3 | 3.5 | 7 | 6.9 | 3.6 | 4.3 | 4.3 | 4.3 |
| | Peak Memory | GB | 9.7 | 12.5 | 18.1 | 18.1 | 18.1 | 19.2 | 3.5 | 3.5 | 3.5 | 4.8 | 7.2 | 7.2 | 3.6 | 4.3 | 4.3 | 4.3 |
| Alignment Metrics | Aligned | # | 11,999,827 ||| | | 11,870,740 | 11,999,763 | 11,999,936 | 11,998,226 | 11,998,804 | 12,000,000 | 11,995,872 | 11,999,975 | 11,999,763 | 11,999,484 | 11,995,422 |
| | | Diff. | | | | | | -129,087 | -64 | 109 | -1,601 | -1,023 | 173 | -3,955 | 148 | -64 | -343 | -4,405 |
| | Properly Paired | # | 11,999,460 ||| | | 11,742,902 | 11,998,912 | 11,999,344 | 11,996,528 | 11,997,254 | 11,999,976 | 11,995,410 | 11,977,218 | 11,998,994 | 11,997,702 | 11,991,992 |
| | | Diff. | | | | | | -256,558 | -548 | -116 | -2,932 | -2,206 | 516 | -4,050 | -22,242 | -466 | -1,758 | -7,468 |
| | Incorrectly Aligned | # | 40,561 ||| | | 138,655 | 143,012 | 141,373 | 147,764 | 85,297 | 95,672 | 99,243 | 58,682 | 150,036 | 15,953 | 61,642 | 61,887 |
| | | Diff. | | | | | | 98,094 | 102,451 | 100,812 | 107,203 | 44,736 | 55,111 | 58,682 | 150,036 | 15,953 | 21,081 | 21,326 |
| | Sensitivity[6] | % | 99.66% ||| | | 97.77% | 98.81% | 98.82% | 98.75% | 99.28% | 99.20% | 99.14% | 99.17% | 99.53% | 99.48% | 99.45% |
| | | Diff. | - ||| | | -1.89% | -0.85% | -0.84% | -0.91% | -0.38% | -0.46% | -0.52% | -0.49% | -0.13% | -0.18% | -0.21% |
| | FDR[7] | % | 0.34% ||| | | 1.17% | 1.19% | 1.18% | 1.23% | 0.71% | 0.80% | 0.83% | 0.83% | 0.47% | 0.51% | 0.52% |
| | | Diff. | - ||| | | 0.83% | 0.85% | 0.84% | 0.89% | 0.37% | 0.46% | 0.49% | 0.49% | 0.13% | 0.18% | 0.18% |

[1] Alignment results by the three entries of SOAP3-dp (1/4 SA, 1/2 SA, Full SA) are identical.
[2] The time consumption of BWA is calculated as "align left reads"+"align right reads"+"sampe". The index loading times of "align right reads" and "sampe" modules are 1 second due to the reason that, index files

were cached during "align left reads". However, datasets larger than the host memory will flush the cache during alignment.

[3] The alignment time consumption of GEM is calculated as "alignment"+"convert to SAM format ". The conversion module was run with 4 threads in consistent with the alignment module.

[4] SOAP3-dp, SOAP3, SeqAlto and GEM aligners explicitly provide index loading time consumption. The index loading time for Bowtie2, CUSHAW2 and BWA are calculated by the total size of index, divided by 100MB/s, which is the average network file system speed of the testing environment. The index loading time maybe underestimated while the time processing the index was not calculated.

[5] The alignment times were explicitly provided by the aligners (include results processing and input/output time) or calculated by total elapsed time minus estimated index loading time.

[6] Sensitivity is calculated as "Correctly aligned reads"/"All simulated reads". The higher the better.

[7] FDR is calculated as "Incorrectly aligned reads"/"All aligned reads". The lower the better.

Table 3. Summary of alignment and variation calling using SOAP3-dp and BWA with different parameters and datasets. PE100 and PE150 represent the 100bp and 150bp paired-end reads of YH sample. 'w/' and 'w/o' indicates whether the alignments were processed with and without GATK's local realignment, respectively. 'dbSNP' is the number of SNPs already archived in dbSNP v135. The running times were rounded to hour.

| Program / Type | Running time (h) | | Reads aligned | Properly Paired | SNP | | | | | | Indels | |
| --- | --- | --- | --- | --- | --- | --- | --- | --- | --- | --- | --- | --- |
| | Aln. | GATK | | | w/o | | | w/ | | | | |
| | | | | | Total | dbSNP | Homozygous | Total | dbSNP | Homozygous | w/o | w/ |
| BWA (Default, PE100) | 99 | 32 | 93.30% | 92.10% | 3,996,730 | 3,926,590 | 1,620,269 | 3,922,841 | 3,876,262 | 1,609,899 | 370,231 | 383,347 |
| BWA (-o 1 -e 50 -L, PE100) | 3363 | 36 | 96.74% | 94.80% | 4,040,813 | 3,955,380 | 1,609,729 | 3,946,627 | 3,896,737 | 1,601,049 | 385,216 | 417,832 |
| SOAP3-dp (Default, PE100) | 6 | 35 | 98.03% | 96.98% | 4,120,165 | 4,028,414 | 1,615,376 | 4,015,725 | 3,956,371 | 1,604,106 | 407,533 | 406,845 |
| BWA (Default, PE150) | 266 | 46 | 90.30% | 89.17% | 3,963,085 | 3,897,866 | 1,621,742 | 3,911,386 | 3,862,977 | 1,616,767 | 350,201 | 384,684 |
| SOAP3-dp (Default, PE150) | 14 | 92 | 97.66% | 96.54% | 4,139,625 | 4,055,314 | 1,612,213 | 4,067,552 | 4,002,750 | 1,612,351 | 409,925 | 422,427 |

# Figure legends

Figure 1, Alignment workflow.
For each read (paired-end specifically, single-end is only with step 1 and step 3), the alignment would be decided in at most three steps. In step 1, SOAP3-dp aligns both ends of a read-pair to the reference genome by using GPU version 2way-BWT algorithm (Methods). Pairs with only one end aligned proceed to step 2 for a GPU accelerated dynamic programming (Methods) alignment at candidate regions inferred from the aligned end. Pairs with both ends unaligned in step 1 and those ends failed in step 2 proceed to step 3 to perform a more comprehensive alignment across the whole genome until all seed hits (substrings from the read) are examined or until a sufficient number of alignments are examined.

Figure 2,
Speed and sensitivity of alignment using simulated paired-end reads. We recorded the number of correct and incorrect alignments stratified by reported mapping quality for each dataset. We then calculated the cumulative number of correct and incorrect alignments from high to low mapping quality. We considered an alignment correct only if the leftmost position was within 50bp of the position assigned by the simulator on the same strand according to the previous study of Bowtie2 to avoid soft-clipping artifacts.

Figure 3,
The accumulated number of incorrectly aligned reads categorized at different mapping quality scores by the five aligners.

Figure 4,
Alignment time consumption of using GPU card "GTX680" and previous generation GPU card "Tesla C2070" respectively.

Figure 5,
**a.** Indel length distribution of indels smaller than or equal to 20bp, **b.** larger than 20bp identified by SOAP3-dp and BWA respectively using full set of 100bp paired-end YH sample reads.

# Methods:

## Implementation details of SOAP3-dp

To align a paired-end read, SOAP3-dp proceeds in three steps (Figure 1). In step 1, SOAP3-dp uses GPU-accelerated 2way-BWT [15] to align those reads without gap opening on to the reference using a 3-level stratified alignment pipeline design. In step 2, for those reads with one end mapped but another end unmapped, a candidate region flanking the mapped end is aligned to the unmapped end using GPU-accelerated dynamic programming algorithm. In step 3, for those reads with both ends unmapped as well as reads still unmapped in step 2, seeds (substrings of a read) are extracted at regular intervals along the read and its reverse complement. SOAP3 module aligns these seeds back to the reference genome and enumerates candidate regions to be aligned to the whole read using dynamic programming.

To better illustrate SOAP3-dp's detailed workflow and parameters, we have prepared two sets of slides, which should be read together with the text.

http://bio8.cs.hku.hk/dataset/Workflow.ppsx
- SOAP3-dp workflow for paired-end alignment
- SOAP3-dp workflow for single-end alignment

http://bio8.cs.hku.hk/dataset/Parameters.ppsx

1. Optimization of parallel access to the GPU global memory.

SOAP3 makes use of the 2way-BWT indexing technique[22] and involves a lot of random access to the indexing data structures in the main memory. The original design of 2way-BWT[5] was based on two-level sampling. The design works well for CPU but not in the highly parallel environment of GPU. The data structures are too large and must be placed into the global memory of GPU, causing serious memory contention among the processors inside the GPU. Therefore, in SOAP3, the index is redesigned to use one-level sampling instead, which greatly reduces the number of memory accesses by half (for details, one may refer to our previous study[23]).

Apart from reducing the number of global memory accesses, we also optimize the time of individual access to the global memory. This is achieved by *coalescing* simultaneous global memory accesses. To illustrate the idea of coalescing, we first need to explain how the GPU handles threads. GPU threads are grouped into units called *warps* for execution on a streaming multiprocessor (SM). The typical size of a warp is 32 threads. At some point, all threads in the same SM access the global memory. Since the threads are working with different data, it is likely that they access different memory locations at the same time. The GPU architecture is designed in a way that, these memory accesses would be much faster if 1) the memory locations accessed are close to each other (e.g. within a 128-byte segment), and 2) no two threads access the same memory location. A group of memory accesses is considered to be *coalesced* if they satisfy the above properties.

In SOAP3-dp, we try to coalesce as many memory accesses as possible. The global memory of GPU device holds two large data structures – the 2way-BWT index, and the set of reads. To enable coalesced access. the set of reads is partitioned into groups of 32 (equals the warp size). For each group, the reads are arranged as follows. Let $w_{i,j}$ denote the $j$-th word of the $i$-th read in the group ($1<=i<=32$). Instead of storing the reads in the most natural way, i.e. $w_{1,1}, w_{1,2}, ... , w_{1,m}, w_{2,1}, w_{2,2}, ... , w_{2,m}, w_{3,1}, ...$ (where $m$ is the number of words occupied by each read), we rearrange them into: $w_{1,1}, w_{2,1}, ... , w_{32,1}, w_{1,2}, w_{2,2}, w_{3,2}, ... , w_{32,2}, w_{1,3}, ...$ . When the threads simultaneously access, say, the first words of the reads, the memory locations accessed are $w_{1,1}, w_{2,1}, ... , w_{32,1}$, forming a contiguous 128-byte segment. These coalesced accesses are done in a single memory transaction, achieving excellent memory throughput. Memory accesses to the BWT index are highly unpredictable, coalescing them is difficult. Nevertheless, the BWT index is designed with all the BWT information for matching a base put together in a memory chunk (64 bytes) that can be assessed using one single memory access.

2. Divergence control and 3-level stratified alignment pipeline.
GPU works in a single-instruction multiple-thread (SIMT) mode. Processors in the same SM must execute the same instruction at one time. When mismatches are allowed, a read can have more than one branch during alignment. Too many diverging branches however would lower the efficiency of GPU drastically, because most processors (with few branches) may become idle and wait for a few others. Therefore, we derive a useful parameter, which can be determined at runtime, whether a read would generate too many branches, and reads are classified into different levels of complexity according to this parameter. The basic idea is that reads of different levels should be aligned separately. In particular, we let the GPU handle the first two levels, and use the CPU to take care of the most complicated reads (which account for a small percentage only). Furthermore, to fully utilize both GPU and CPU processing power, SOAP3 overlaps the alignment of complicated reads from the previous batch in CPU with the alignment of the next batch in GPU (as shown in the Supplementary Figure 4).

3. GPU-accelerated dynamic programming.
To perform dynamic programming for aligning a read with a candidate region in genome, Smith-Waterman algorithm is applied. However, a straightforward implementation of the algorithm does not fit well under the GPU environment due to the large number of memory accesses. Therefore another implementation is suggested so that the number of memory accesses can be reduced by half. In the following, the straightforward implementation is first described, and then it is shown how to be modified in order to decrease the number of memory accesses. Similar implementations have also been applied in ClustalW[24] and CUDASW++ before[25].

Given a candidate region $T$ (of length $m$) *and* a read $R$ (of length $n$), the aim is to find a sub-region $T'$ inside $T$ such that the alignment score between $R$ and $T'$ is maximum.

1 Let $M(i,j)$ be the maximum alignment score between all suffixes of $T[1...i]$ and
2 $R[1...j]$. The resulting score would be: $max_{1\leq i\leq m} M(i,n)$.
3
4 Let $S_{MA}$, be the score for match, and let $S_{MI}$, $S_{GO}$, $S_{GE}$ be the penalty scores for
5 mismatch, gap opening and gap extension. It is also needed to define $I(i,j)$ as the
6 maximum alignment score all suffixes of $T[1...i]$ and between $R[1...j]$ under the
7 condition that $R[j]$ is aligned to a space, and $D(i,j)$ as the maximum alignment
8 score between all suffixes of $T[1...i]$ and $R[1...j]$ under the condition that $T[i]$ is
9 aligned to a space.
10
11 The recursive formulas are as follows:

$$I(i,j) = \max \begin{cases} M(i,j-1) + S_{GO} \\ I(i,j-1) + S_{GE} \end{cases}$$

$$D(i,j) = \max \begin{cases} M(i-1,j) + S_{GO} \\ D(i-1,j) + S_{GE} \end{cases}$$

$$M(i,j) = \max \begin{cases} M(i-1,j-1) + \delta(R[i],T[j]) \\ I(i,j), D(i,j) \end{cases}$$

12 where $\delta(x,y) = S_{MA}$ if $x = y$, or $\delta(x,y) = S_{MI}$ if $x \neq y$.
13
14 And the base cases are:

$$I(i,0) = -\infty, i = 1...m$$
$$D(0,j) = -\infty, j = 1...n$$

15 $M(i,0) = 0, i = 0...m$

$$M(0,j) = S_{GO} + (j-1)S_{GE}, j = 1...n$$

16
17 A straightforward approach of the implementation is as follows:
18

---

Smith-Waterman algorithm: Compute $I$, $D$ and $M$

1: Initialize tables $I$, $D$ and $M$ according to the base cases.
2: For $i = 1 \rightarrow m$ do
3:  For $j = 1 \rightarrow n$ do
4:   $I(i,j) \leftarrow \max\{M(i,j-1) + S_{GO}, I(i,j-1) + S_{GE}\}$
5:   $D(i,j) \leftarrow \max\{M(i-1,j) + S_{GO}, D(i-1,j) + S_{GE}\}$
6:   $M(i,j) \leftarrow \max\{M(i-1,j-1) + \delta(R[i],T[j]), I(i,j), D(i,j)\}$
7:  End for
8: End for

---

19
20 To implement this straightforward approach on GPU, tables $I$, $D$ and $M$ are
21 created inside the GPU's global memory. The fact that the access of GPU's global
22 memory is much slower than its arithmetic operation affects the efficiency of the
23 algorithm. In every loop, there are 7 table-reading and 3 table-writing operations.
24
25 It is realized that backtracking could be done with only tables $D$ and $M$, thus the
26 table $I$ can be eliminated. Therefore, another approach is suggested as follows,
27 which requires only 2 table-reading and 2 table-writing operations in each loop.
28

---

Improved implementation of Smith-Waterman algorithm for GPU:
Compute $D$ and $M$

1: Declare register variables: $M^u$, $M^d$, $V_M$, $V_I$, $V_D$

---

    (note: $M^u$ refers to $M(i\text{-}1, j\text{-}1)$ and $M^d$ refers to $M(i\text{-}1, j)$)
2:  Initialize tables $D$ and $M$ according to the base cases.
3:  For $i = 1 \to m$ do
4:    $$M^d \leftarrow 0, V_M \leftarrow 0, V_I \leftarrow -\infty$$
5:    For $j = 1 \to n$ do
6:      $$M^u \leftarrow M^d$$
7:      $$M^d \leftarrow M(i-1, j)$$
8:      $$V_I \leftarrow \max\{V_M + S_{GO}, V_I + S_{GE}\}$$
9:      $$V_D \leftarrow \max\{M^d + S_{GO}, D(i-1, j) + S_{GE}\}$$
10:     $$V_M \leftarrow \max\{M^u + \delta(R[i], T[j]), V_I, V_D\}$$
11:     $$D(i, j) \leftarrow V_D$$
12:     $$M(i, j) \leftarrow V_M$$
13:    End for
14:  End for

4. Effort Limit for Dynamic Programming.

Reads with seeds that match an exceedingly large amount of places on the genome can spur an excessively large number of dynamic programming problems. For example, a poly-A homopolymer could match over ten thousand loci in the genome. SOAP3-dp avoids executing an excessive number of dynamic programming problems by adopting a ceiling on the number of candidate regions in step 2 and step 3. Candidate regions are scored with number of supporting seeds and sorted descendingly. If the ceiling is set to 30, for example, SOAP3-dp will only perform dynamic programming alignment in the best 30 candidate regions. The ceiling is set in the configuration file, but values higher than the default may strongly affect the performance with limited accuracy improvement.

5. Paired-end alignment.

SOAP3-dp supports alignment of paired-end reads in which both ends of a single DNA fragment are sequenced. The user sets expected minimum and maximum fragment lengths using –v and –u parameters, as well as orientations of the ends in configuration file (typically, Illumina uses Forward-Reverse while SOLiD uses Forward-Forward). A paired-end alignment that matches these expectations is called "properly paired" and an alignment that violates these expectations is "unpaired". If a pair fails to be aligned as properly paired, SOAP3-dp attempts to align each end individually. This is similar to both BWA's and Bowtie2's behavior. When a read pair fails to be aligned properly but both ends could be aligned individually, SOAP3-dp reports these alignments.

In contrast to BWA and SOAP2, which rely on a mapped end to determine a candidate region for further dynamic programming alignment, SOAP3-dp could align those reads with both ends unmapped. This allows read pairs from large period of variation hotspots to be aligned.

6. Scoring functions.
Details discussed in Supplementary Note.

### Simulation of single-end and paired-end reads:

Mason 0.1[20] was used to simulate reads using the GRCh37 major build human reference genome, including 22 pairs of autosomes, 2 sex chromosomes and a mitochondrial chromosome . For the paired Illumina-style datasets with read length ranging from 50bp to 250bp, Mason was run in 'Illumina' read mode with options -N 6000000 --source-no-N -mp -sq -ll 500 -le 25 -rn 2 -hn 2 --haplotype-snp-rate 0.001 --haplotype-indel-rate 0.0001 --haplotype-no-N -n 100 -pi 0 -pd 0 --no-N'. Each PE set was simulated to contain exactly 12 million reads (6M pairs), which is about 2 times the default batch size of SOAP3-dp.

All simulated datasets are available at http://bio8.cs.hku.hk/dataset/.

### Simulation comparison between tools:

Executable files for SOAP3-dp v2.3, Bowtie2 v2.0.0-beta4, BWA 0.6.2, SeqAlto basic 0.5-r123, BarraCUDA_r232, CUSHAW-1.0.40, CUSHAW2-v2.1.9, SOAP3_version146 and GEM-core_i3-20121106-022124 were obtained via standard build procedures with default arguments. We indexed the reference genome with each tool's default indexing parameters. SeqAlto uses 22bp seed length and sub-sampled mode. "Running time" was measured from initial call of the aligner to the completion of SAM-format output. 'Reads aligned' was measured as the number of reads for which the tool found at least one alignment regardless of mapping score. 'Properly paired' was measured as the number of read-pairs aligned with proper read orientation and insert-size range (mean insert-size ± 3*standard deviation). 'Peak memory' and 'Average memory' usage was measured by tracking the Linux's *proc* file-system with respective *process id*. For BWA and BarraCUDA, separate calls of the software modules were required for aligning each end and for processing intermediate alignment results into a final SAM file. 'Running time' was measured for separate modules while 'Peak memory' and 'Average memory' were measured across all modules. For GEM, to be consistent with other tools, a conversion is necessary after alignment to obtain SAM format results with mapping quality. Time consumptions were measured separately and then summed for comparison to other tools. All tools or components were run with 4 threads (except for the alignment module of BarraCUDA, where the CPU thread is constantly 1). Parameters were listed in Supplementary Note as receipts.

The experiments used a single computing node running CentOS v6.3 with an Intel i7-3930k 3.2Ghz quad-core processor, an Nvidia GTX 680 GPU card with 4GB non-ECC (Error-correcting code) graphic memory and 64GB non-ECC memory.

Scripts and command lines to evaluate the authenticity of aligned reads and generate the ROC curves are available at http://bio8.cs.hku.hk/dataset/.

### YH data production:

Genomic DNA was isolated using standard molecular biology techniques. For each short insert library, 5 μg of DNA was fragmented, end-repaired, A-tailed and ligated to Illumina paired-end adapters. The ligated fragments of 100bp paired-end reads (PE100) were size-selected at 170bp and 500bp on agarose gels, while PE150 were size-selected at 240bp. All libraries are amplified by LM-PCR to yield the corresponding short insert libraries. PE100 were sequenced using TruSeq v2 while PE150 were sequenced using TruSeq v3 reagent on the Illumina sequencing platform.

**Real data comparison:**
The 100bp and 150bp paired-end Illumina HiSeq 2000 reads of YH sample were sequenced and deposited to EBI with study accession number ERP001652. The data are also available at http://yh.genomics.org.cn.

SOAP3-dp uses default parameters. BWA uses both default parameters and "-o 1 –e 50", which allows at most a gap not longer than 50bp (-m option to elevate the 2M hits limit for each read was not applied due to out of memory error, the option allows more reads to be aligned but consumes much more memory and longer alignment time). The latter option allows more reads to be aligned and more indel signals to be discovered, but would enormously decrease the running speed. Alignments were post-processed by following procedures: 1) local realignment by GATK v2.1, 2) duplication removal by Picard v1.74, 3) base quality score recalibration, 4) variants calling by UnifiedGenotyper and 5) variants quality score recalibration by GATK v1.6. Step 1 is optional according to the experiment while steps 2 to 5 are mandatory. Parameters and known variant sets were set according to the GATK's Best Practice v4 on GATK's website.

SOAP3-dp used a single node with a quad-core Intel Xeon E5570 2.93Ghz CPU and a GPU while BWA used 10 nodes with the same CPU. To imitate the real production environment, we used Nvidia Tesla C2070 GPU device with 6G graphic memory and with ECC enabled to perform full YH dataset alignment.

**Fosmid sequencing:**
Fosmid libraries (averagely 40kbp in size) were constructed according to Kim et al. [26]. In total, ~100k Fosmid clones were created and every 30 Fosmids were pooled together sharing a barcode for Illumina HiSeq 2000 sequencing. For each pool, one 200bp and one 500 bp insert size libraries were constructed and sequenced at 20x respectively. If a library had problem of abnormal base content bias or a relatively high base error rate reported by base-calling software, we took it as a non-qualified library and performed sequencing again. Each pool was assembled with SOAPdenovo[27] using 63-mer and other parameters as default. Sequences solved by SOAPdenovo's repeat solving module were remembered and soft-masked in final sequences in order not to obscure the following alignment. The assembled sequences were aligned to the human reference genome using BWASW with default parameters. Most of the Fosmid clones could be assembled to full length. For fragmented Fosmid clone sequences, we further assembled the fragments according to the in-pool linkage information during alignment. We require over 90% of a Fosmid clone sequence to be linearly aligned to only one location in the reference genome. Finally, 460 Fosmid clones were found covering the 50 randomly selected SOAP3-dp specific deletion calls that are not yet archived in dbSNP v135. We define "a Fosmid sequence supporting a deletion" as over 80% of the deleted bases in reference genome cannot be covered by the aligned Fosmid sequence (excluding soft-masked bases), and the identity of the 200bp alignments flanking the deletion should exceed 90%. While a Fosmid clone can only come from a haploid, we require a heterozygous deletion has at least a Fosmid supporting the deletion, while a homozygous deletion should only have Fosmids supporting the deletion. Heterozygous deletions with lower than 5 spanning Fosmids and without a Fosmid supporting the deletion will be classified as "not clear". Homozygous deletions without a spanning Fosmid will also be classified as "not clear" (Supplementary Tables 15, 16).

The assembled Fosmid sequences are available as Supplementary Data in BAM file format. Raw reads are available upon request.

**Experiment on Amazon EC2:**
The experiment used a single GPU Quadruple Extra Large Instance (cg1.4xlarge, $2.1 per hour) rented from the Amazon Elastic Compute Cloud (EC2) service (http://aws.amazon.com/ec2). The instance has 2 quad-core Intel Xeon X5570 at 2.93GHz with hyper-threading, 2 Nvidia Tesla M2050 GPU cards with 3GB ECC graphic-memory per card, 22 gigabytes of physical memory and runs Amazon Linux AMI v2012.09 operating system. Alignments were distributed onto the two GPU cards with two SOAP3-dp processes sharing the same copy of index in host memory. Each process occupies at most 7 threads. 10 sets of Illumina HiSeq 2000 reads generated in 1000 genomes project (Supplementary Note) were downloaded from the Amazon Simple Storage Service (S3). Additional tests have been carried out by NIH biowulf laboratory (http://biowulf.nih.gov/apps/bioinf-gpu.html) and Tianhe-1A super-computing center (http://en.wikipedia.org/wiki/Tianhe-I). While these clouds host CPU and GPU computing nodes and centralized storage system, the ultra-fast SOAP3-dp could be easily integrated into existing pipelines.

# Reference:

1. (2010) A map of human genome variation from population-scale sequencing. Nature 467: 1061-1073.
2. Li H, Ruan J, Durbin R (2008) Mapping short DNA sequencing reads and calling variants using mapping quality scores. Genome Res 18: 1851-1858.
3. Langmead B, Trapnell C, Pop M, Salzberg SL (2009) Ultrafast and memory-efficient alignment of short DNA sequences to the human genome. Genome biology 10: R25.
4. Li H, Durbin R (2009) Fast and accurate short read alignment with Burrows-Wheeler transform. Bioinformatics 25: 1754-1760.
5. Li R, Yu C, Li Y, Lam TW, Yiu SM, et al. (2009) SOAP2: an improved ultrafast tool for short read alignment. Bioinformatics 25: 1966-1967.
6. Li Y, Zheng H, Luo R, Wu H, Zhu H, et al. (2011) Structural variation in two human genomes mapped at single-nucleotide resolution by whole genome de novo assembly. Nat Biotechnol 29: 723-730.
7. Mullaney JM, Mills RE, Pittard WS, Devine SE (2010) Small insertions and deletions (INDELs) in human genomes. Hum Mol Genet 19: R131-136.
8. Mu JC, Jiang H, Kiani A, Mohiyuddin M, Bani Asadi N, et al. (2012) Fast and accurate read alignment for resequencing. Bioinformatics 28: 2366-2373.
9. Liu Y, Schmidt B (2012) Long read alignment based on maximal exact match seeds. Bioinformatics 28: i318-i324.
10. Marco-Sola S, Sammeth M, Guigo R, Ribeca P (2012) The GEM mapper: fast, accurate and versatile alignment by filtration. Nat Methods 9: 1185-1188.
11. Zaharia M, Bolosky WJ, Curtis K, Fox A, Patterson D, et al. (2011) Faster and More Accurate Sequence Alignment with SNAP. arXiv 1111.5572v1.
12. Ewing B, Green P (1998) Base-calling of automated sequencer traces using phred. II. Error probabilities. Genome Res 8: 186-194.
13. Nvidia (2012) Bioinformatics and Life Sciences.
14. Klus P, Lam S, Lyberg D, Cheung MS, Pullan G, et al. (2012) BarraCUDA - a fast short read sequence aligner using graphics processing units. BMC Res Notes 5: 27.
15. Liu CM, Wong T, Wu E, Luo R, Yiu SM, et al. (2012) SOAP3: ultra-fast GPU-based parallel alignment tool for short reads. Bioinformatics 28: 878-879.
16. Wang J, Wang W, Li R, Li Y, Tian G, et al. (2008) The diploid genome sequence of an Asian individual. Nature 456: 60-65.
17. Li H, Durbin R (2010) Fast and accurate long-read alignment with Burrows-Wheeler transform. Bioinformatics 26: 589-595.
18. McKenna A, Hanna M, Banks E, Sivachenko A, Cibulskis K, et al. (2010) The Genome Analysis Toolkit: a MapReduce framework for analyzing next-generation DNA sequencing data. Genome Res 20: 1297-1303.
19. Li H, Handsaker B, Wysoker A, Fennell T, Ruan J, et al. (2009) The Sequence Alignment/Map format and SAMtools. Bioinformatics 25: 2078-2079.
20. Doring A, Weese D, Rausch T, Reinert K (2008) SeqAn an efficient, generic C++ library for sequence analysis. BMC Bioinformatics 9: 11.
21. DePristo MA, Banks E, Poplin R, Garimella KV, Maguire JR, et al. (2011) A framework for variation discovery and genotyping using next-generation DNA sequencing data. Nat Genet 43: 491-498.

22. Lam TW, Li R, Tam A, Wong S, Wu E, et al. High throughput short read alignment via bi-directional BWT; 2009. IEEE. pp. 31-36.
23. Liu C, Lam T, Wong T, Wu E, Yiu S, et al. (2011) SOAP3: GPU-based compressed indexing and ultra-fast parallel alignment of short reads. The 3rd Workshop on Massive Data Algorithmics (MASSIVE 2011). Paris, France.
24. Larkin MA, Blackshields G, Brown NP, Chenna R, McGettigan PA, et al. (2007) Clustal W and Clustal X version 2.0. Bioinformatics 23: 2947-2948.
25. Liu Y, Maskell DL, Schmidt B (2009) CUDASW++: optimizing Smith-Waterman sequence database searches for CUDA-enabled graphics processing units. BMC Res Notes 2: 73.
26. Kim UJ, Shizuya H, de Jong PJ, Birren B, Simon MI (1992) Stable propagation of cosmid sized human DNA inserts in an F factor based vector. Nucleic Acids Res 20: 1083-1085.
27. Li R, Zhu H, Ruan J, Qian W, Fang X, et al. (2010) De novo assembly of human genomes with massively parallel short read sequencing. Genome Res 20: 265-272.

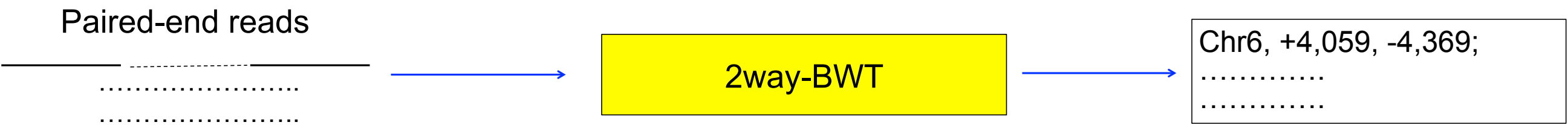
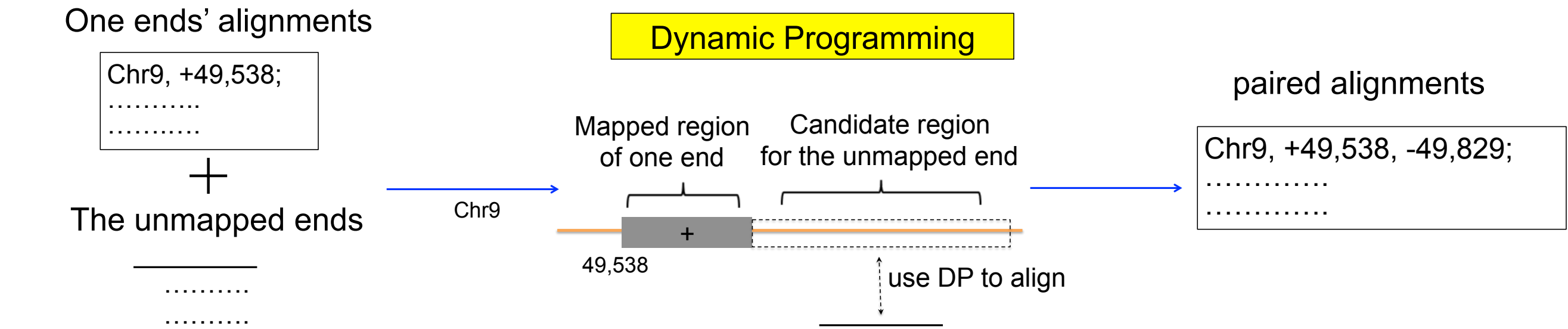
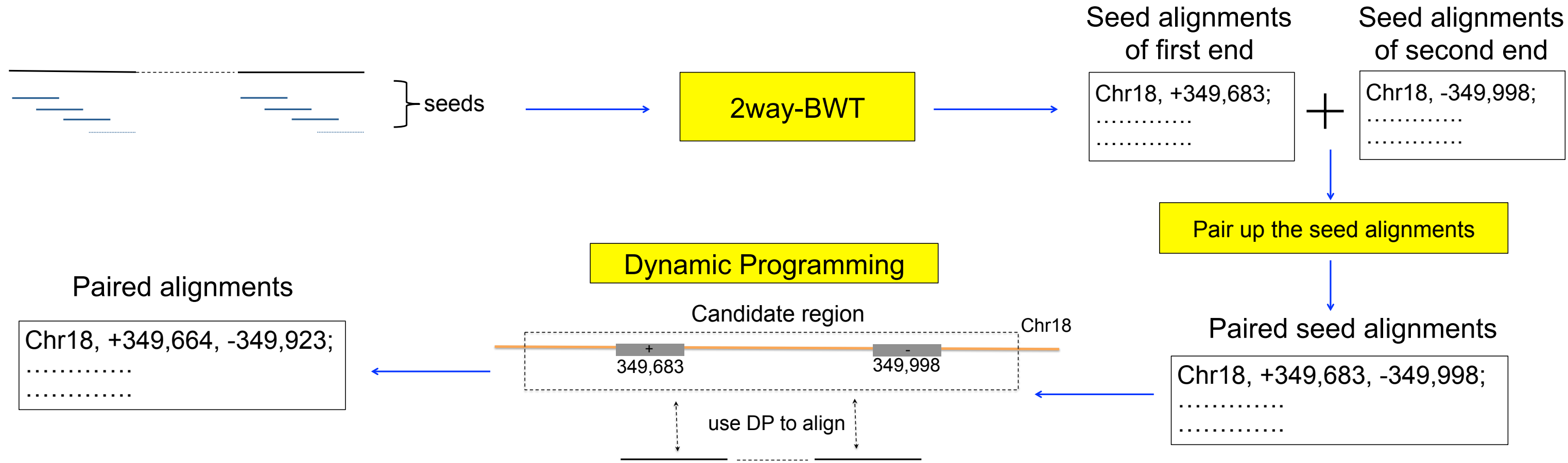

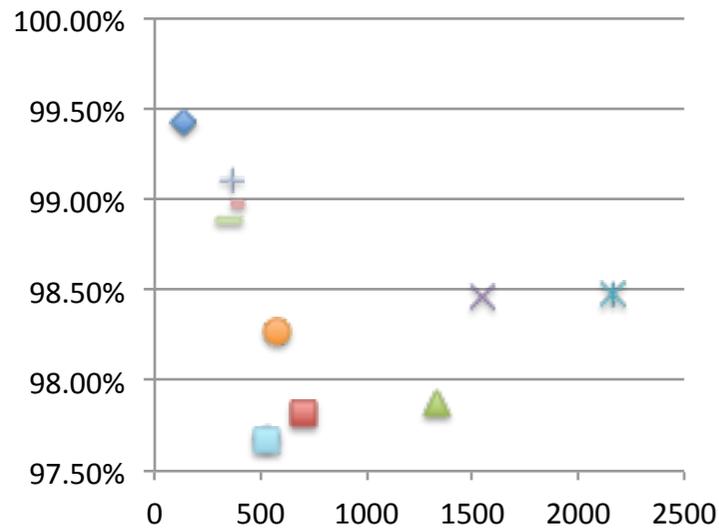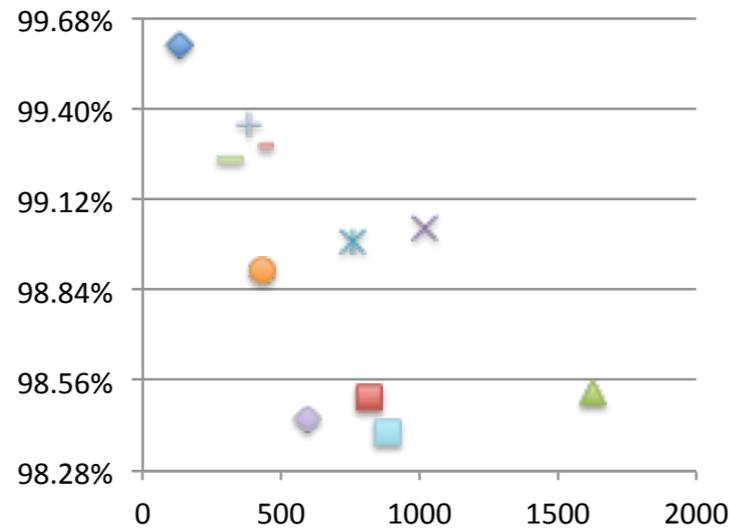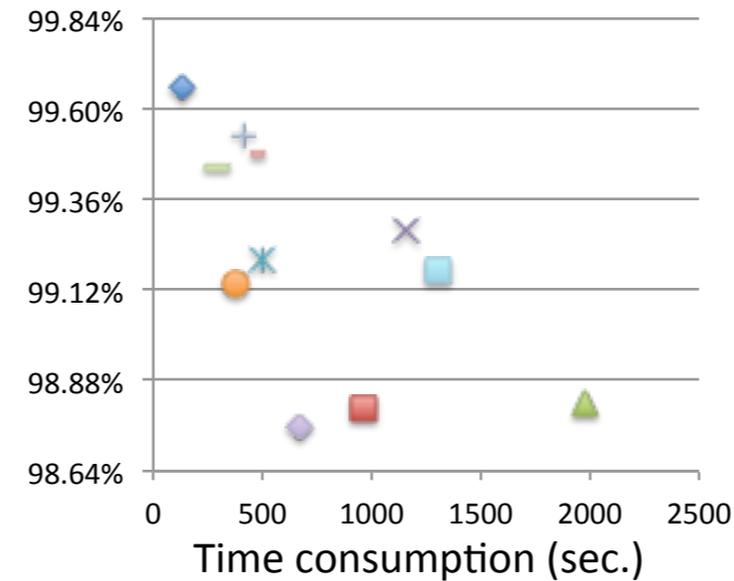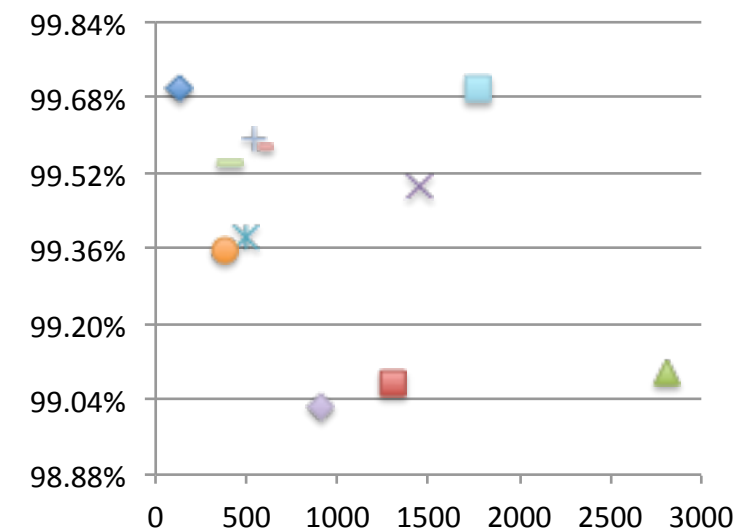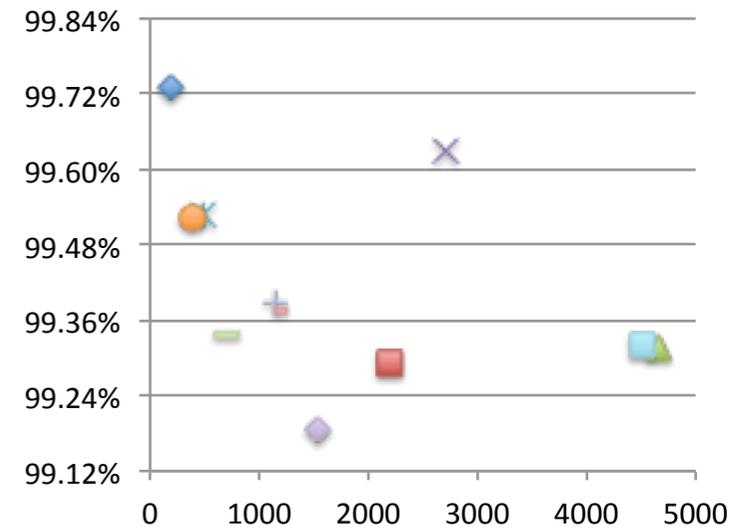
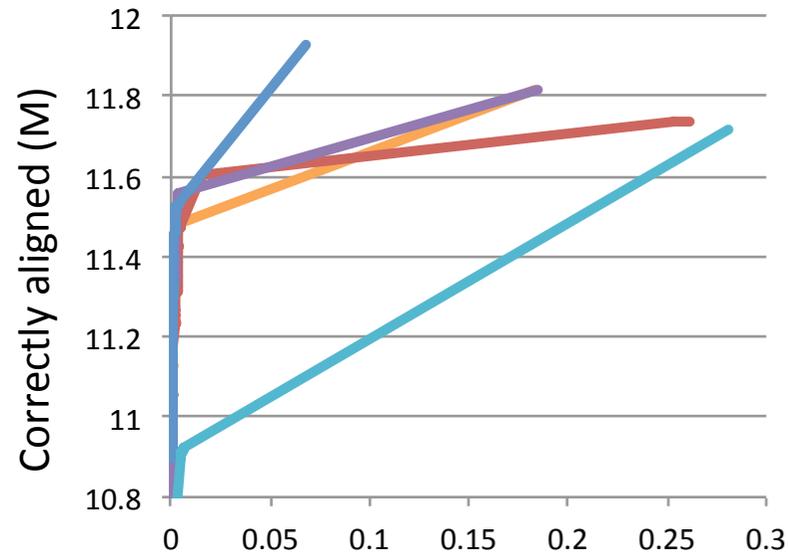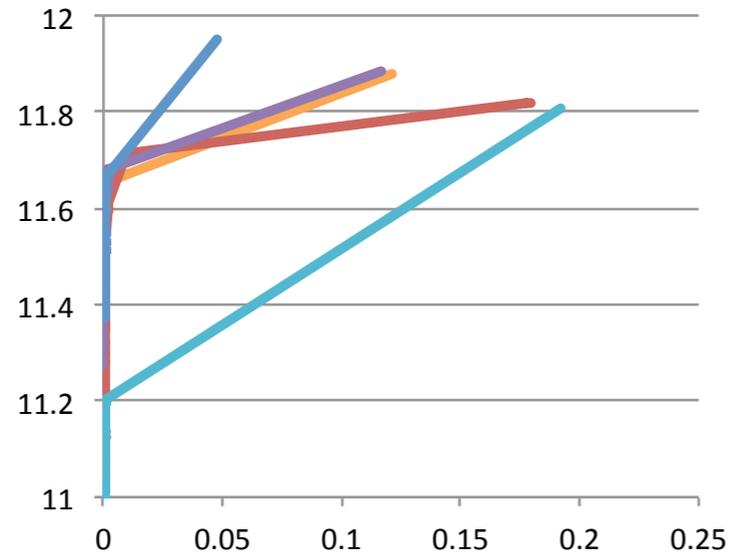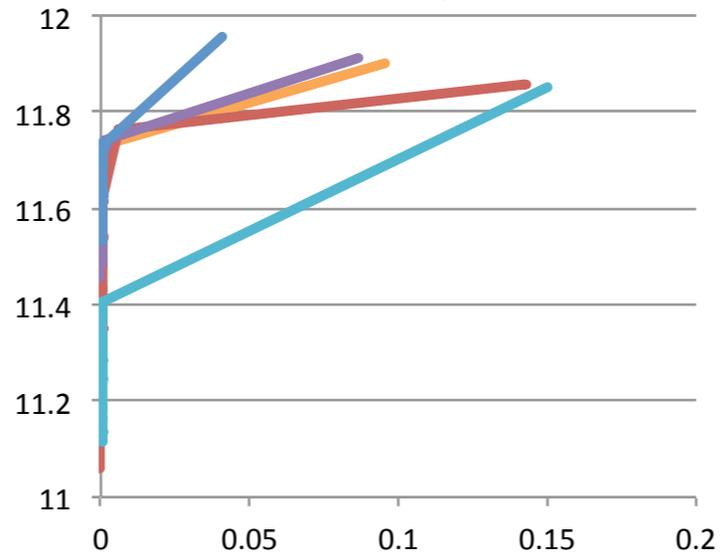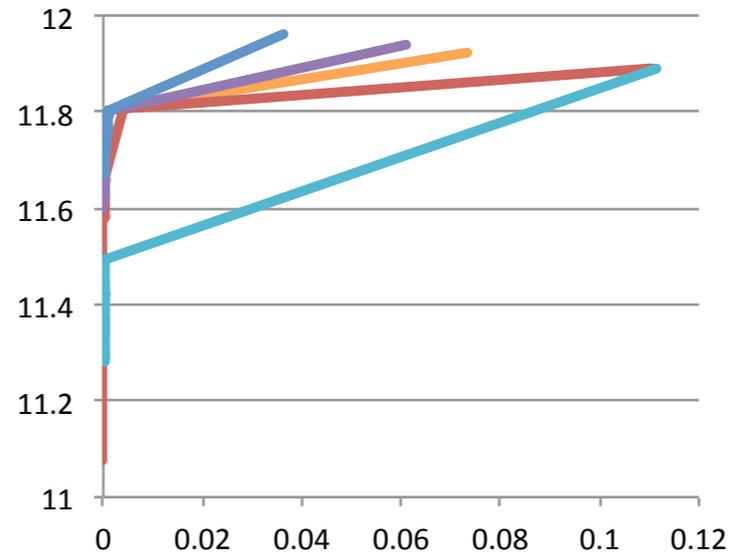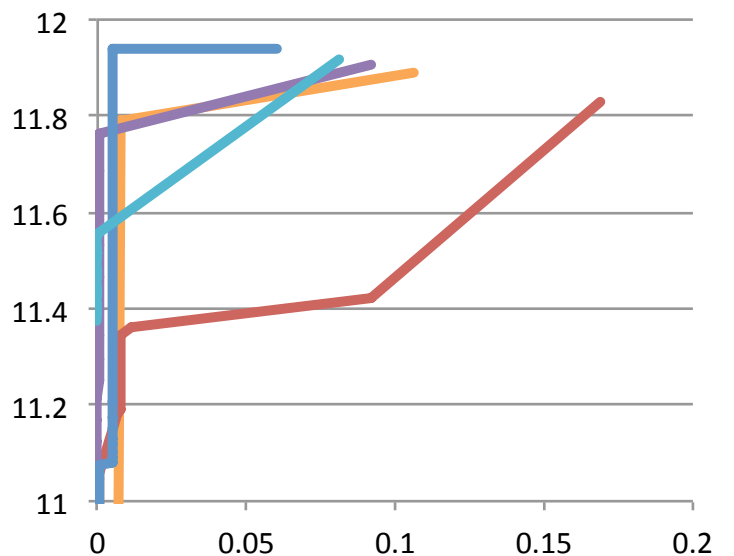
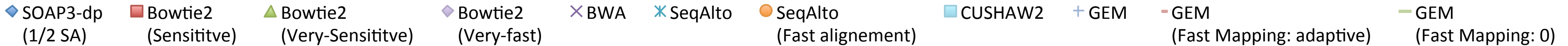

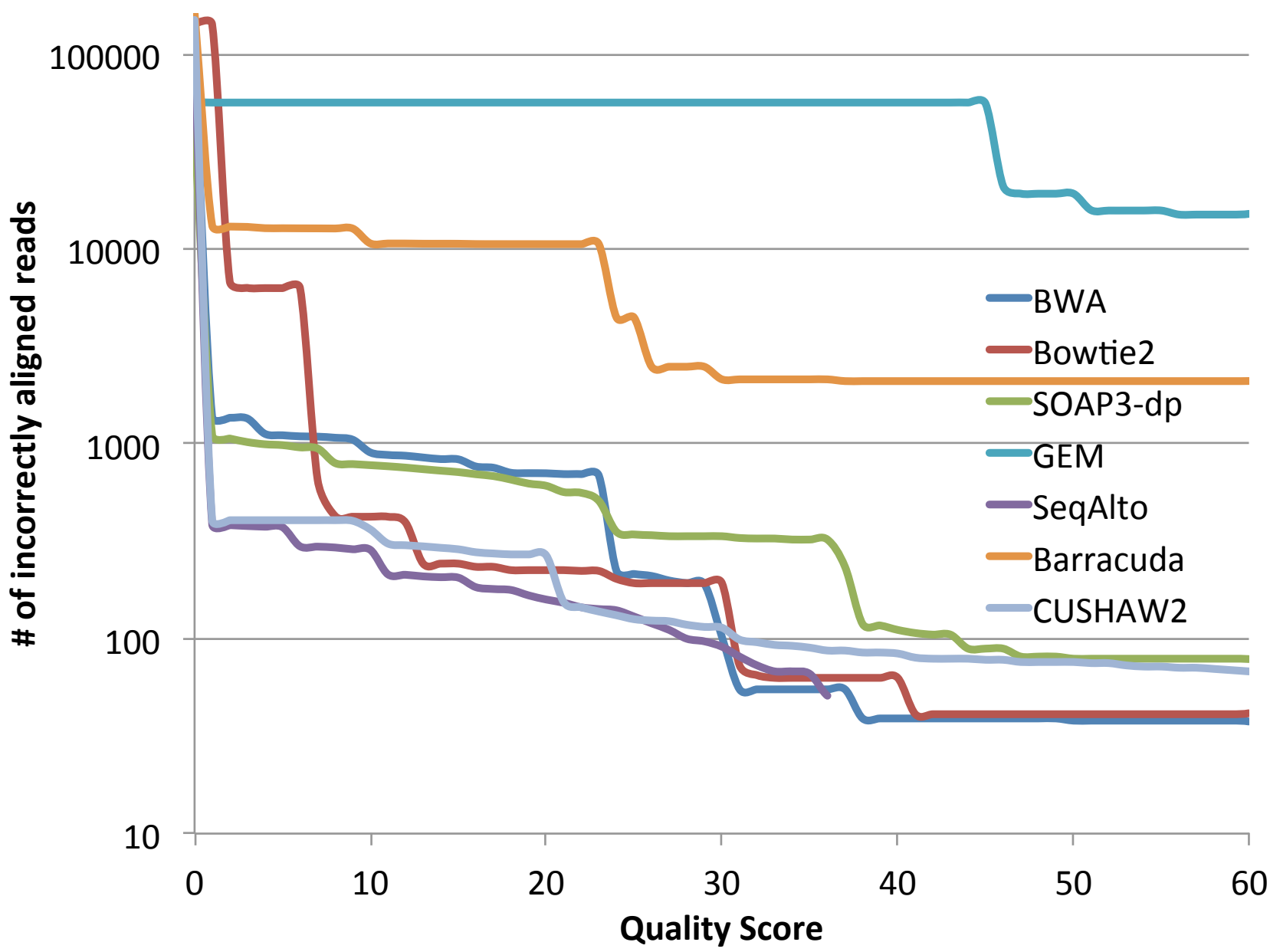

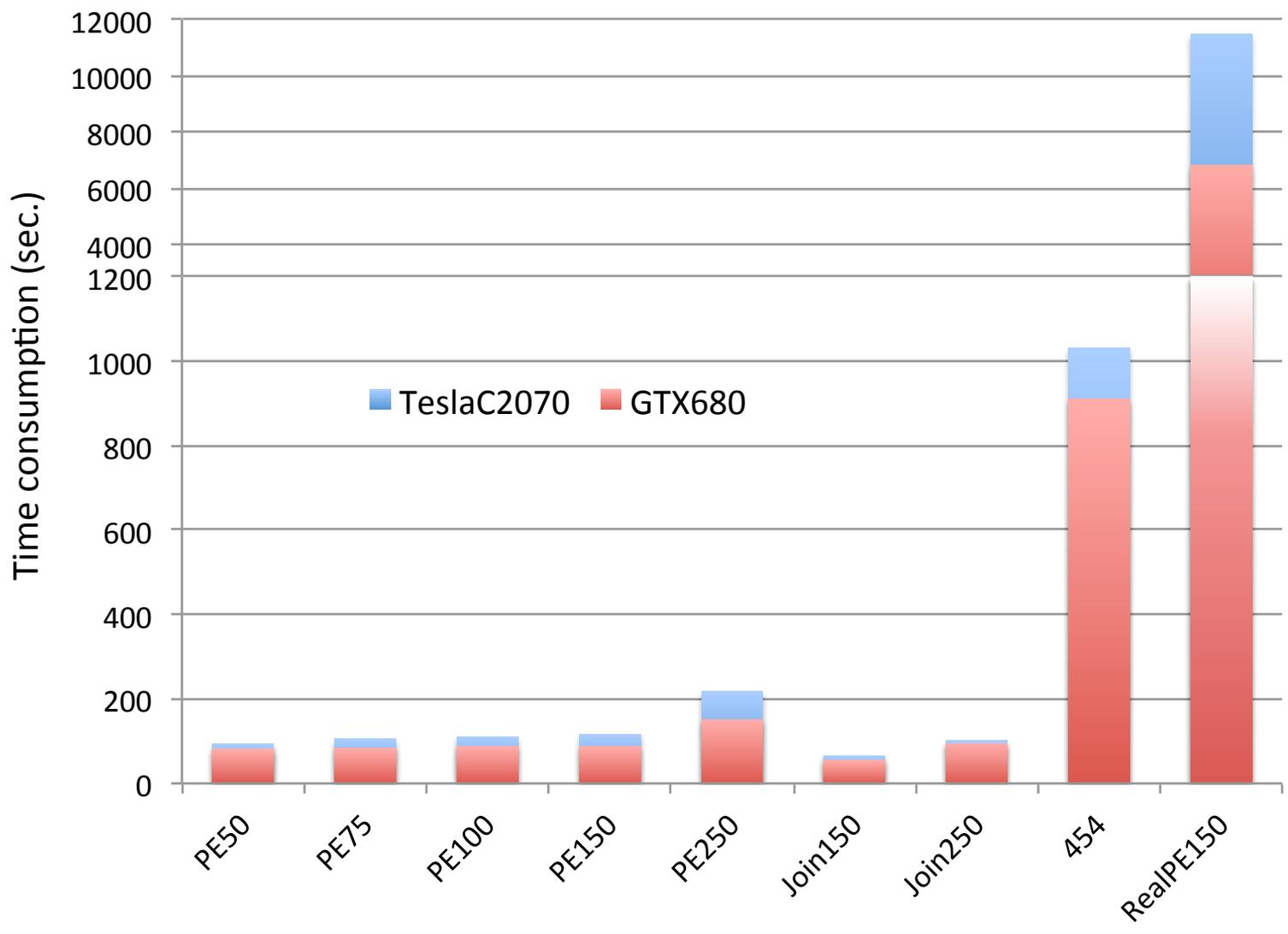

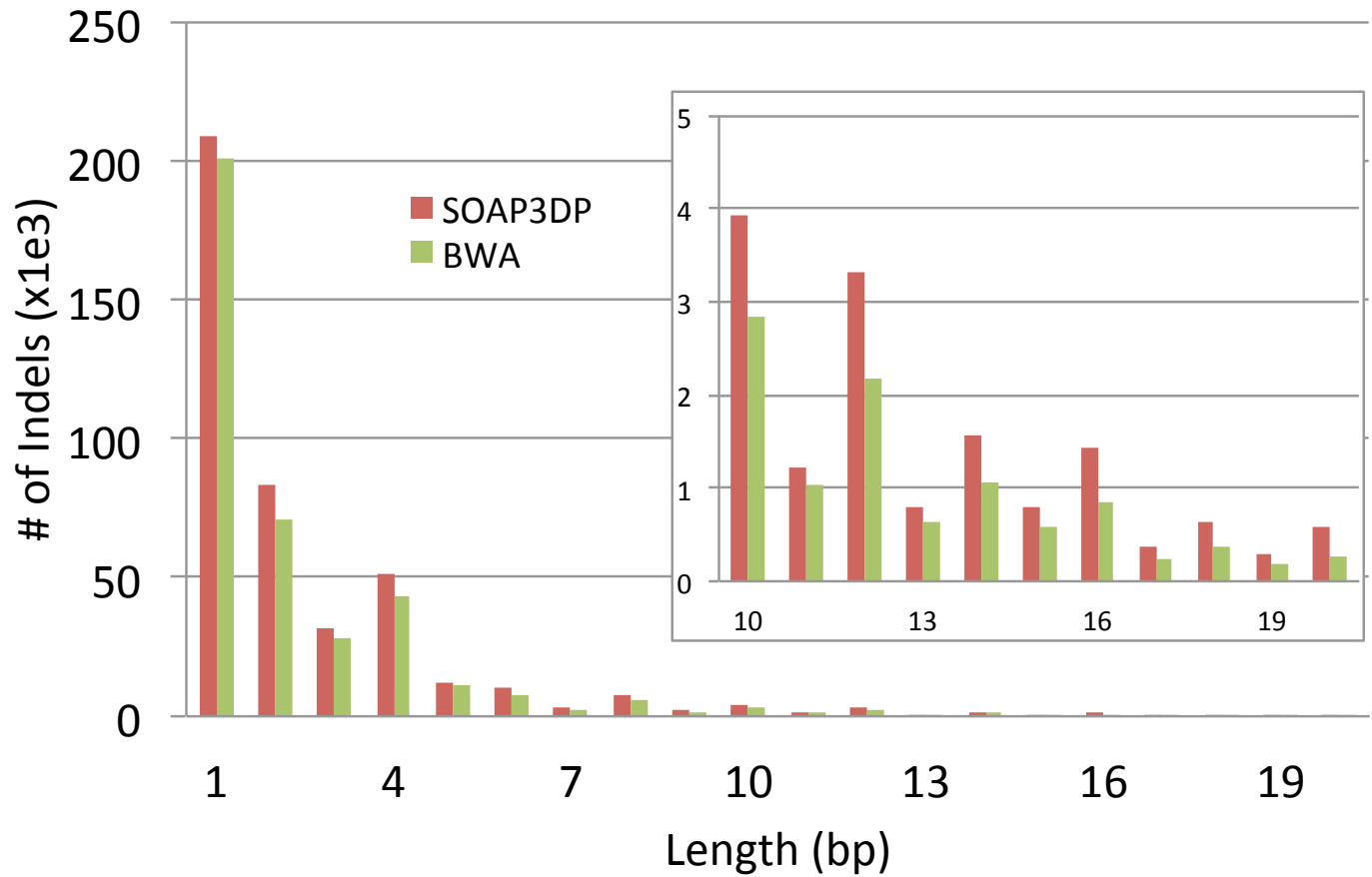

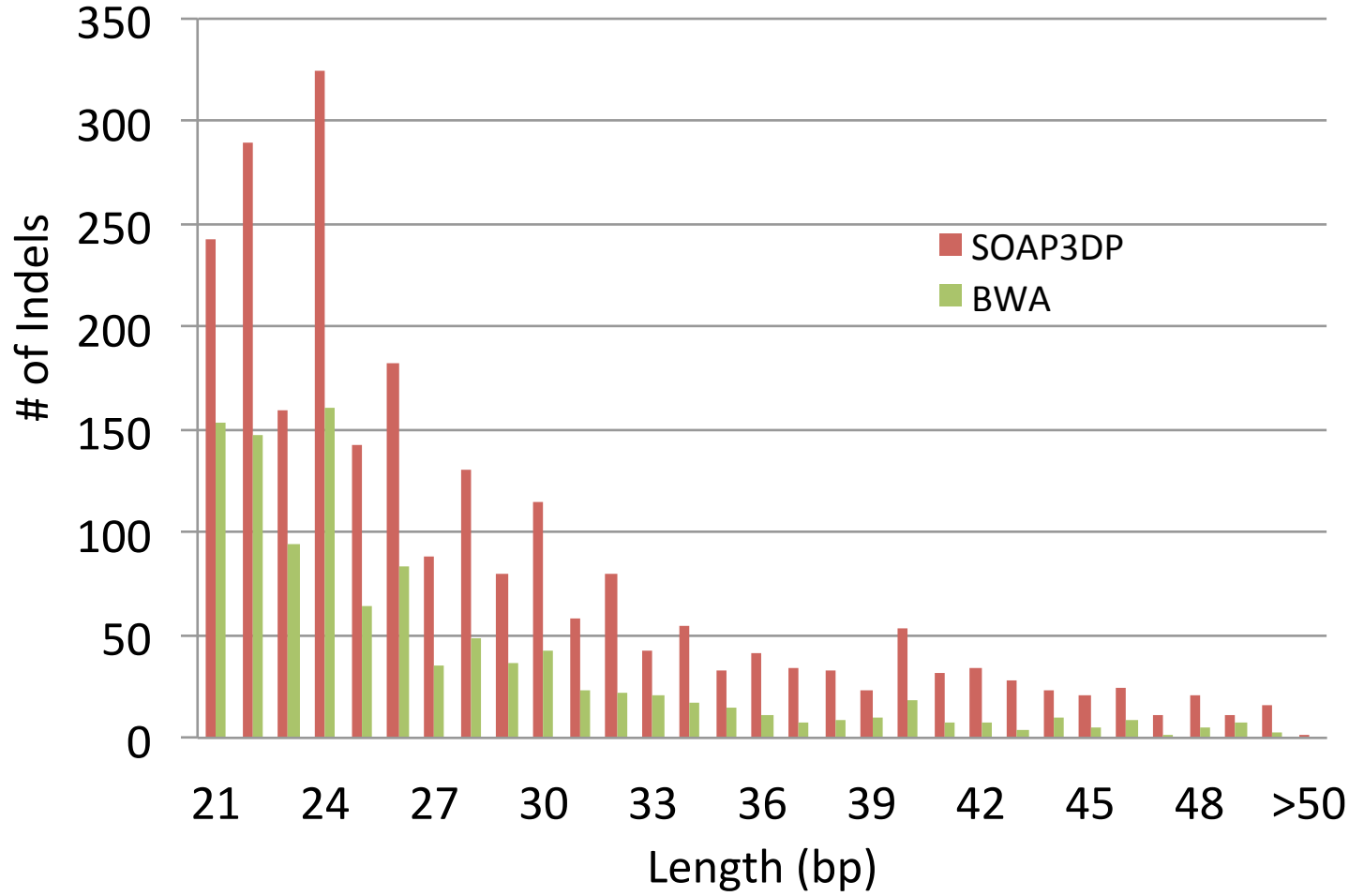

Indel length distribution (>20bp)